\documentclass[%
 reprint,
 amsmath,amssymb,
 aps,
]{revtex4-1}

\usepackage{graphicx}
\usepackage{dcolumn}
\usepackage{bm}
\usepackage{longtable}
\usepackage{comment}

\usepackage{color, colortbl}
 \usepackage{graphicx}
\usepackage{rotate}
\definecolor{Gray}{gray}{0.85}
\newcolumntype{a}{>{\columncolor{Gray}}c}

\begin{document}

\preprint{APS/123-QED}

\title{
Lattice vibrations and electronic properties  of GaSe nanosheets from first principles }

\author{Mousa Bejani}
\affiliation{Dept. of condensed matter, Faculty of Physics, University of Tabriz, 29 Bahman Blvd., 5166616471 Tabriz, Iran}
\affiliation{Dept. of Physics, and INFN, University of Rome Tor Vergata, Via della Ricerca Scientifica 1, I-00133 Rome, Italy}

\author{Olivia Pulci}
 \email{olivia.pulci@roma2.infn.it}
\affiliation{Dept. of Physics, and INFN, University of Rome Tor Vergata, Via della Ricerca Scientifica 1, I-00133 Rome, Italy}

\author{Jamal Barvestani} 
\author{Ali Soltani Vala}
\affiliation{Dept. of condensed matter Physics, Faculty of Physics, University of Tabriz, 29 Bahman Blvd., 5166616471 Tabriz, Iran}

\author{Friedhelm Bechstedt}

\affiliation{IFTO, Friedrich Schiller Universit\"at, Max-Wien Platz 1, Jena, Germany
}%

\author{Elena Cannuccia}
\affiliation{Universit\'e Aix-Marseille, Laboratoire de Physique des Interactions Ioniques et Moléculaires (PIIM), UMR CNRS 7345, F-13397 Marseille, France}
\affiliation{Dept. of Physics, and INFN, University of Rome Tor Vergata, Via della Ricerca Scientifica 1, I-00133 Rome, Italy}

\date{\today}

\begin{abstract}
Electronic properties and lattice dynamics of bulk $\varepsilon$-GaSe and one, two and three tetralayers GaSe are investigated by means of density functional and density functional perturbation theory.
The few-tetralayers systems are semiconductors with an indirect nature of the fundamental band gap and a Mexican-hat-shape is observed at the top of the valence band. The phonon branches analysis reveals the dynamical stability for all systems considered together with the LO-TO splitting breakdown in two-dimensional systems. In-plane (E) and out-of-plane (A) zone-center lattice vibrations dominate the Raman and IR spectra. 
\end{abstract}

\pacs{Valid PACS appear here}
\maketitle


\section{\label{sec:level1}Introduction}
The discovery of graphene \cite{Geim.Novoselov:2007:NM} with its unique physical and chemical properties has triggered the research on other two-dimensional (2D) inorganic materials, like transition metal dichalcogenides and transition metal oxides \cite{Miro.Audiffred.ea:2014:CSR,Niu.Li:2015:PiSS}. Several three-dimensional (3D) van der Waals (vdW) solids have been shown to be exfoliable down to atomic monolayers (MLs), giving rise to fabrication of several 2D crystals \cite{Koski.Cui:2013:AN}. Stable single layer or few layers of 2D compounds can be produced via chemical exfoliation or mechanical cleavage starting from the corresponding III-VI layered crystal counterparts \cite{Ashton.Paul.ea:2017:PRL,Bandurin.Tyurnina.ea:2016:NN,Hu.Wen.ea:2012:AN}. In parallel, the interesting research field of heterostructures fabrication by stacking different 2D crystals on top of each other has flourished \cite{Geim.Grigorieva:2013:N,Cannuccia2012,Prete2017,Prete2018}. 

Among the transition-metal monochalcogenides MX with M = Ga, In and X = S, Se, Te, GaSe is a prototypical example. It consists of covalently bonded Ga and Se atoms, vertically stacked in four-atom Se-Ga-Ga-Se layers (here on named tetralayers, TL) held together by vdW forces. The single tetralayer thickness is 0.93~nm \cite{Hu.Wen.ea:2012:AN}. There are four possible stacking arrangements for bulk GaSe, designated as $\beta$, $\varepsilon$, $\gamma$, and $\delta$ \cite{Kuhn.Chevy.ea:1975:pssa}. The $\beta$, $\varepsilon$ polytypes have two-tetralayer hexagonal stacking sequences (2H), i.e., eight atoms, in the unit cell, and they crystallize respectively into $P_{3}/mmc$ and $P\overline{6}m2$ space groups. The $\delta$ polytype is 4H, while the $\gamma$ polytype crystallizes in a rhombic or trigonal lattice (3R). With a band gap of about 2~eV \cite{LeToullec.Balkanski.ea:1975:PLA} and a strong optical oscillator strength, bulk GaSe polytypes have potential optoelectronic applications in the visible spectral  range. Electronic confinement effects in GaSe nanosheets may shift the optical absorption edge toward higher energies and increase the oscillator strength \cite{Lei.Ge.ea:2013:NL}.

The $\varepsilon$-GaSe is among the polytypes the most common and the most studied one. 
First-principles calculations \cite{Rybkovskiy.Arutyunyan.ea:2011:PRB,Ma.Dai.ea:2013:PCCP} predict the valence band maximum (VBM) to be situated at $\Gamma$ point, and the conduction band minimum (CBM) close to point M of the hexagonal Brillouin zone (BZ) whereas the direct gap is only slightly different from the indirect one.  
Similar investigations have been performed for GaSe layered structures with respect to the slab thickness~\cite{Rybkovskiy.Arutyunyan.ea:2011:PRB,Ma.Dai.ea:2013:PCCP,Zolyomi.Drummond.ea:2013:PRB,Zhuang.Hennig:2013:CM,Li.Li:2015:NR} with the inclusion in some cases of the quasi-particle effects within Hedin's GW approximation \cite{Rybkovskiy.Arutyunyan.ea:2011:PRB,Zhuang.Hennig:2013:CM}. The increase of the fundamental gap with a decreasing thickness of the slab has been demonstrated. It is accompanied by the formation of a Mexican-hat or camel-back dispersion of the uppermost valence band near the $\Gamma$ point, making the few-tetralayer systems indirect semiconductors.

In contrast to the electronic structure, the phonons of GaSe are less investigated. Since Raman spectroscopy is employed to determine the thickness (and thus, the number of unit layers) of ultrathin flakes in a non-destructive, and unambiguous manner, Raman studies are available now for bulk and a-few-tetralayer systems \cite{Lei.Ge.ea:2013:NL,Ibragimov1989,Late.Liu.ea:2012:AFM,Zhou.Nie.ea:2014:AN,Rodriguez.Mueller.ea:2014:JoVS&TB,Rahaman2018}.
Theoretical investigations of phonon modes have been only performed for a tetralayer of GaSe \cite{Zolyomi.Drummond.ea:2013:PRB,Pandey.Parker.ea:2017:N}. For that reason, in this paper we investigate the vibrational properties of $\varepsilon$-GaSe and layered GaSe by varying the slab thickness from bulk, 3 TL down to 1 TL.  
The methods are described in Sec.~\ref{sec2}. In section~\ref{sec3} the results on electronic and dynamical properties are discussed. Finally, a summary and conclusions are given in Sec.~\ref{sec4}.

\section{\label{sec2}Computational methods}

The \emph{ab initio} calculations are based on Density Functional Theory (DFT) using different exchange-correlation (xc) functionals as implemented in the {\tt Quantum Espresso} (QE) package \cite{quantum:espresso}. The electron-ion interaction is described by pseudopotentials for Ga and Se cores. Thereby the semicore Ga 3$d$ electrons are treated as valence electrons. Both the local density approximation (LDA) \cite{Kohn.Sham:1965} and the generalized gradient approximation (GGA) as proposed by Perdew, Burke, and Ernzerhof (PBE) \cite{Perdew.Burke.ea:1996} are applied. The influence of the vdW interaction between the covalently bonded Se-Ga-Ga-Se tetralayers  is studied by adding a semiempirical  dispersion (D) potential to the DFT total energy, through a pairwise force field according to Grimme's DFT-D2 method \cite{Grimme.Antony.ea:2010:TJoCP}.

\begin{figure}[h]
\includegraphics[scale=0.35]{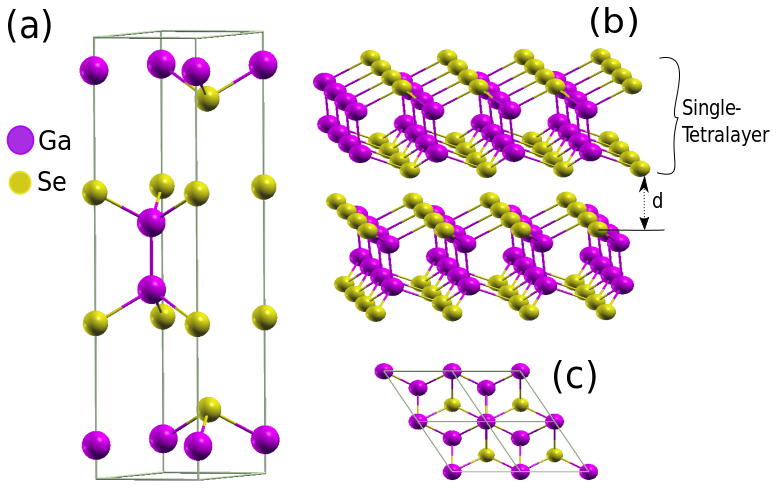}
\caption{\label{fig1} Side (a) and top view (c) of the $\varepsilon$-GaSe bulk unit cell, (b) GaSe bulk and single tetralayer perspective view. $d$ is the interlayer distance.}
\end{figure}

For all considered GaSe structures the calculations of electronic properties are performed using periodic boundary conditions and plane-wave expansion of the electronic eigenfunctions. The bulk unit cell is represented in Fig.~\ref{fig1} with its side and top view. The nanosheets with two and three tetralayers are stacked using the single tetralayer (see Fig.~\ref{fig1} (b)) as building block and respecting the $z$-stacking of the bulk \cite{Bechstedt:2003:Book:PSP}. A vacuum spacing of 28~{\AA} between the periodic images of mono-, bi-, and tri-tetralayers is chosen in order to diminish the electronic interaction between the nanosheets. To sample the Brillouin Zone (BZ) in the total energy and self-consistent electronic structure calculations, a Monkhorst-Pack (MP) special point mesh \cite{Monkhorst.Pack:1976:PRB} of 12$\times$12$\times$1 (6$\times$6$\times$1, 6$\times$6$\times$1, 6$\times$6$\times$6) is applied in the case of a GaSe 1TL, (2TL, 3TL, bulk $\varepsilon$ polytype). The structural optimizations are performed by relaxing the atomic geometries till forces are less than $10^{-6}$ a.u. The plane-wave energy cutoff and the total energy threshold are set to 300~Ry and $10^{-8}$~Ry, respectively. All DFT approaches lead to a significant underestimation of fundamental energy gaps \cite{Bechstedt:2015:Book}. Since GW self-energy calculations would be too expensive, we investigate the band structures of the GaSe geometries within an approximated quasiparticle approach. Explicitly, we apply the hybrid xc-functional HSE06 \cite{Heyd.Scuseria.ea:2003}.

Dynamical properties, i.e., the phonon dispersions as well as Raman frequencies, are calculated within Density Functional Perturbation Theory (DFPT) \cite{Baroni.Gironcoli.ea:2001:RMP} as implemented in the QE package. In particular, we use the {\tt  qe-6.3} version because it implements a truncated Coulomb interaction in the direction parallel to the layered-system stacking orientation ($z$-direction for structures periodic in the $x$-$y$ plane). The implementation of the cutoff guarantees the correct treatment of the LO-TO splitting breakdown in polar two-dimensional materials~\cite{Sohier-2Dcutoff}. 

As a compromise for a reasonable simulation of strong intrasheet and weak intersheet interactions we also apply here the LDA xc-functional~\cite{Kohn.Sham:1965}. A plane-wave cutoff of 90~Ry and robust convergence criteria (total energy/forces for the ionic minimization and 
self-consistence criteria) are necessary to get the correct interatomic force constants to be used in the phonons calculations. A mesh of 21$\times$21$\times$1 k-points is used for the BZ integration and sampling of the reciprocal space for 1TL-GaSe. For 2TL and 3TL cases, given the increasing number of atoms and the volume of the simulation cell, we adopted a less dense q-grid (12$\times$12$\times$1) to keep the calculations reasonable in terms of cpu time. The length of the unit cell along the $z$ direction was chosen to be larger than twice the thickness of the slab  (including electrons) \cite{Sohier-2Dcutoff}. For bulk $\varepsilon$-GaSe we used a mesh of 18$\times$18$\times$6 k-points and 9$\times$9$\times$3 mesh for the sampling of the reciprocal space. 

\section{\label{sec3}Results and Discussion}
\subsection{\label{sec3a}Atomic geometries}

\begin{table*}[t]
\caption{Parameters of atomic geometry (in {\AA}) such as lattice constants $a$, $c$ and atomic distances. Electronic band gaps (in eV)  are also listed.}
\centering
\begin{tabular}{|c|c|c|c|c|c|c|c|c|c|} \hline
xc-functional & $a$ & $c$ & Se-Se   & Ga-Ga   & Ga-Se   & Se-Se   & Indirect & Direct   & Fundamental \\
         &     &     & (intra) & (intra) & (inter) & (inter) & band gap & band gap & gap (in HSE) \\
\hline
bulk     &     &     &         &         &         &         &          &          & \\
LDA-D2   & 3.651 & 15.336 & 4.727 & 2.370 & 2.415 & 3.618 &   & 0.97 &       \\
LDA      & 3.72  & 15.56  & 4.73  & 2.41  & 2.44  & 3.73  &   & 0.97 &       \\
GGA-D2   & 3.662 & 15.587 & 4.812 & 2.432 & 2.470 & 3.826 &   & 0.92 & 1.75 \\
Exp. \cite{Kuhn.Chevy.ea:1975:pssa,Li.Li:2015:NR} & 3.735 & 15.887 & 4.766 & 2.383 & 2.485 & -- &   & 2.12 & \\ \hline
trilayer &     &     &         &         &         &         &          &          & \\
LDA-D2   & 3.704 &        & 4.749 & 2.409 & 2.437 & 3.804 & 1.30 & 1.34 &       \\
LDA      & 3.71 & & 4.74 & 2.41 & 2.44 & 3.82 & 1.30 & 1.34 & \\
GGA-D2   & 3.753 &        & 4.834 & 2.446 & 2.472 & 3.835 & 1.31 & 1.35 & 2.25 \\ \hline
bilayer  &     &     &         &         &         &         &          &          & \\
LDA-D2   & 3.704 &        & 4.742 & 2.405 & 2.437 & 3.812 & 1.54 & 1.63 &  \\
LDA      & 3.71 & & 4.73 & 2.41 & 2.44 & 3.75 & 1.54 & 1.63 & \\
GGA-D2   & 3.748 &        & 4.818 & 2.428 & 2.472 & 3.845 & 1.60 & 1.67 & 2.58 \\\hline
monolayer &     &     &         &         &         &         &          &          & \\
LDA      & 3.73  &    & 4.72    & 2.41  &      &      & 2.16 & 2.30   & \\
GGA-D2   & 3.745 &    & 4.821   & 2.429 &      &      & 2.20 & 2.32 & 3.24 \\ \hline
\end{tabular}
\label{tab1}
\end{table*}

The results of the atomic geometry optimization of bulk GaSe and few-tetralayer systems are summarized in Table~\ref{tab1}. Bulk $\varepsilon$-GaSe (hexagonal space group $P\bar{6}m2$ $(D^1_{3h})$ with point group $\bar{6}m2$ $(D_{3h})$) contains in the unit cell two Se-Ga-Ga-Se layers rotated against each other (Fig.~\ref{fig1}a). The nanosheets with one and three tetralayers crystallize to the same space group as the bulk $\varepsilon$-GaSe, while the 2TL exhibits a reduced symmetry. It belongs to the trigonal space group $P3m1(C^1_{3v})$ with the point group $3m$ $(C_{3v})$.

The $a$ and $c$ lattice constants of $\varepsilon$-GaSe are underestimated with respect to the experimental values independently of the xc-functional employed. Surprisingly, the LDA bulk lattice constants are on the average the closest ones to the experimental data. Differences are less than 0.01~{\AA}. For what concerns the dependence of in-plane lattice constants with the layer thickness we observe that with the vdW functionals, the $a$ value tends to increase. An homogeneity of the $a$ values with respect to the xc-functional is found in few-tetralayer systems. Going from LDA-D2 to LDA and finally to GGA-D2 the intra- and interatomic distances exhibit the average trend to increase. However, for a given xc-functional the trend with the number of layers is not unique. For LDA a decrease of both intralayer and interatomic distances with the decrease of the system thickness is observed. With vdW interaction the intratomic distances decrease whereas the Se-Se interlayer distances increase with decreasing material thickness. Independently of the xc-functional the Ga-Se interlayer distances remain the same. The found atomic geometries are reasonably well described also in LDA, in agreement with the literature \cite{Rybkovskiy.Arutyunyan.ea:2011:PRB,Zolyomi.Drummond.ea:2013:PRB}.

\subsection{\label{sec3b}Electronic structures}

\begin{figure*}[t]
\includegraphics[scale=0.75]{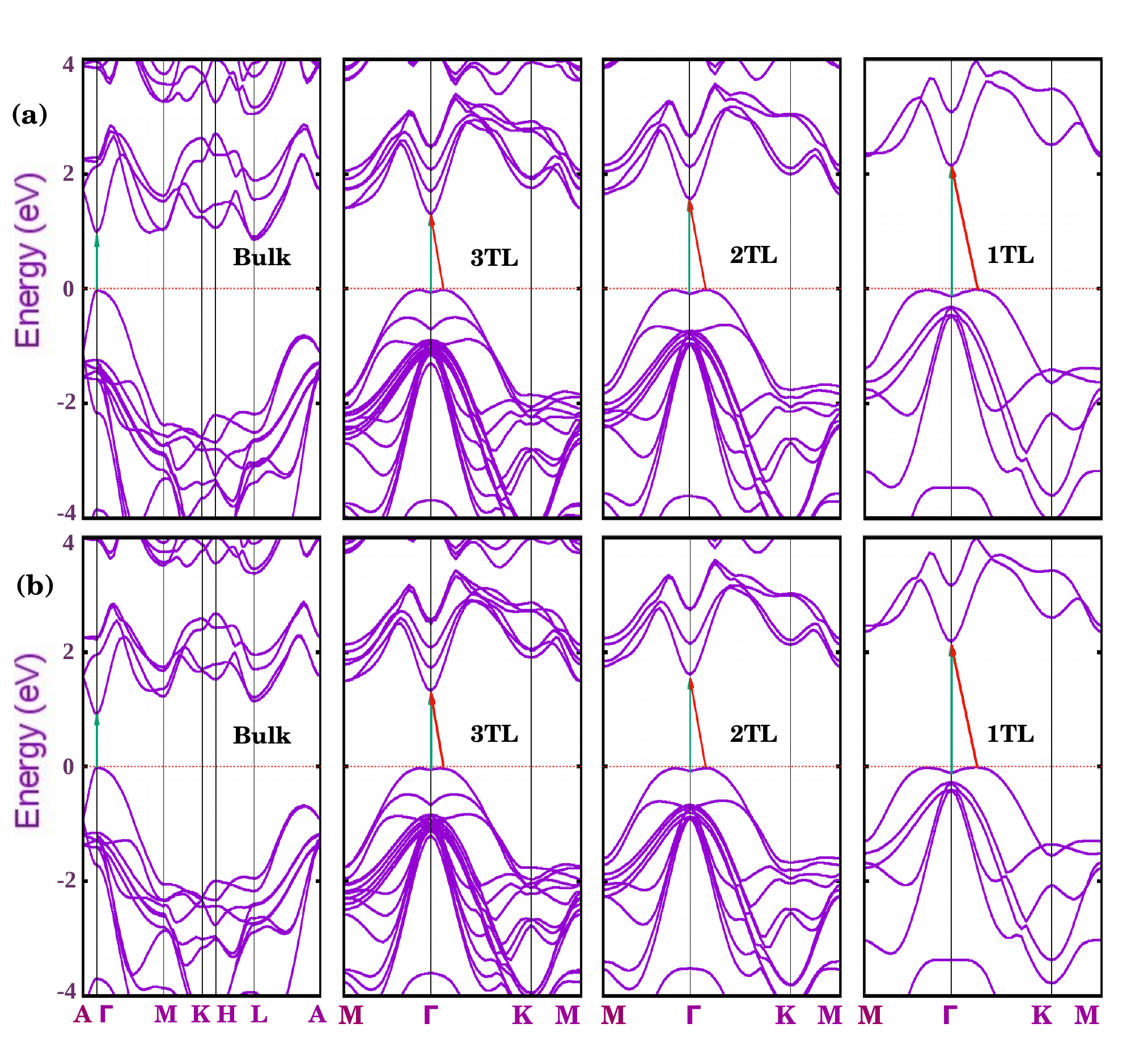}
\caption{\label{fig2}Electronic band structure of bulk $\varepsilon$-GaSe, 3TL, 2TL, and 1TL GaSe as derived from Kohn-Sham eigenvalues calculated by means of two different xc-functionals, (a) GGA-PBE and (b) LDA. The bands are plotted versus high-symmetry lines $M-\Gamma-K-M$ in the 2D hexagonal BZ. In the bulk case also lines A$\Gamma$ and KH along the hexagonal axis and HL and LA on the boundary of the 3D hexagonal BZ are included. Direct (indirect) gaps are illustrated by blue (red) lines. }
\end{figure*}

In order to illustrate the electronic structure of the GaSe systems,  we display their band structures  in Fig.~\ref{fig2} for two different xc-functionals, GGA-PBE and LDA. Despite the neglect of quasiparticle corrections, all the studied systems represent semiconductors, whose gaps increase from bulk to a single tetralayer by more than 50~\% (see Table~\ref{tab1}) because of the electronic confinement effects. Within the HSE06 functional the gap gets increased by about 0.9~eV (also see Table~\ref{tab1}), independent of the confinement. Comparing with the measured gap of $\varepsilon$-GaSe a gap increase of about 1.15~eV should be expected instead. A quasiparticle gap at $\Gamma$ of about 2.34 eV for bulk GaSe, and of  $3.89$~eV for a GaSe 1TL, are predicted  within  GW calculations \cite{Rybkovskiy.Arutyunyan.ea:2011:PRB}. The optical gaps, however, may be significantly smaller especially in the 1TL case, due to the large binding energies of the  excitons in 2D systems\cite{Pulci.Marsili.ea:2015:PSSB,Pulci2012}. This is in agreement with the value of 2.58 eV experimentally obtained from optical-absorption spectroscopy \cite{Rybkovskiy.Arutyunyan.ea:2011:PRB}. A larger optical gap of 3.3~eV as derived from the photocurrent spectrum has been explained by optical transitions from the second valence band to the conduction band minimum \cite{Lei.Ge.ea:2013:NL}.

Bulk $\varepsilon$-GaSe comes out in our calculations  to be a direct gap semiconductor with the lowest optical transitions at the $\Gamma$ point between parabolic conduction and valence bands (see Fig.~\ref{fig2}). The top of the valence band at $\Gamma$ possesses different masses along and perpendicular to the hexagonal axis. All the few-tetralayer systems are indirect semiconductors with gap values in agreement with previous calculations \cite{Rybkovskiy.Arutyunyan.ea:2011:PRB,Ma.Dai.ea:2013:PCCP,Zolyomi.Drummond.ea:2013:PRB}. The uppermost valence band forms a Mexican-hat or camel-back dispersion near the $\Gamma$ point. As a consequence of the strong Rashba contribution to the spin-orbit interaction  the indirect $\Gamma$K $\rightarrow$ $\Gamma$ gap is smaller than the direct $\Gamma$ $\rightarrow$ $\Gamma$ gap (see Table~\ref{tab1}).

\subsection{\label{sec3c}Phonon dispersions}

\begin{figure*}[t]
\centering
\begin{tabular}{cc}
\includegraphics[scale=0.3]{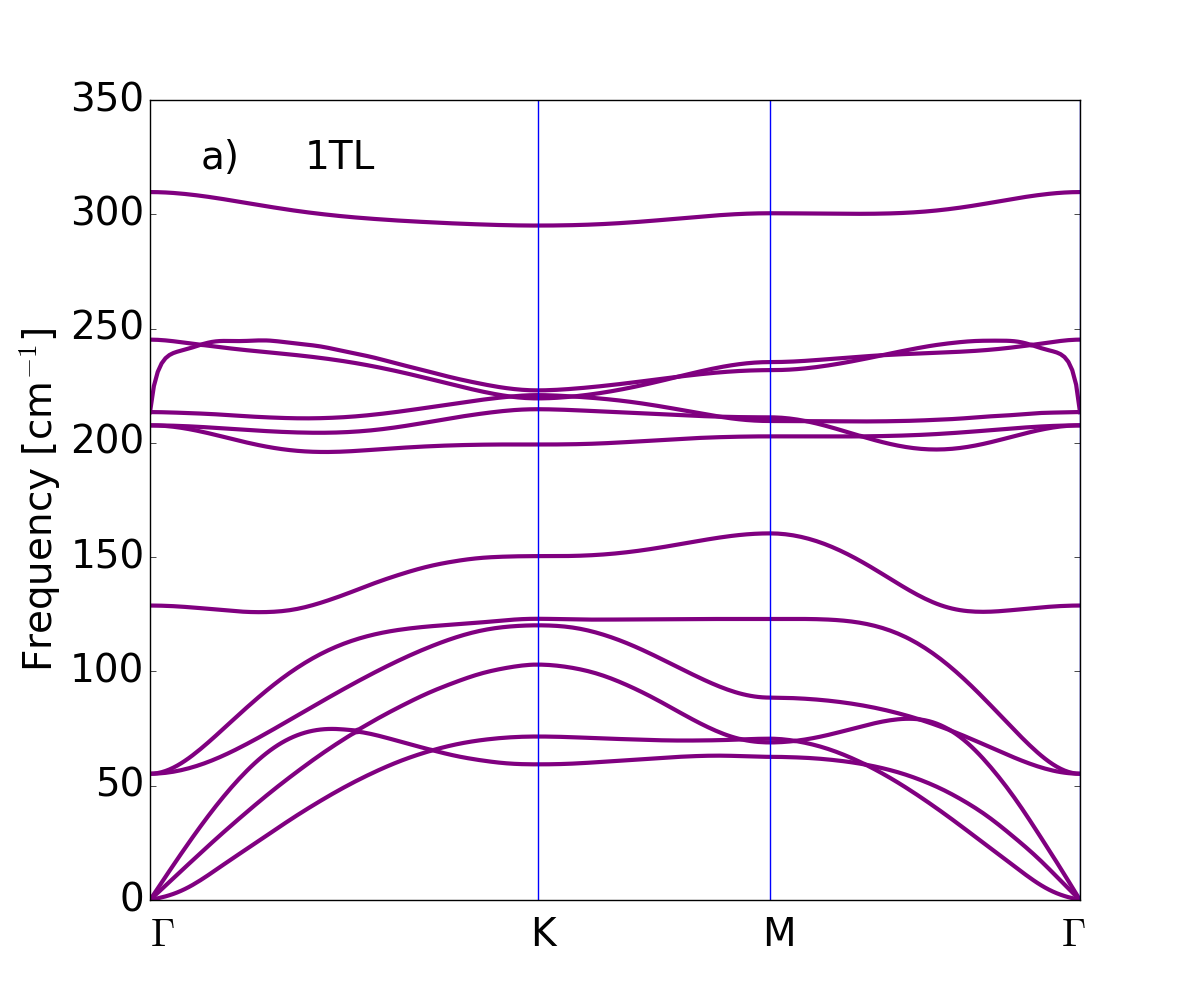}&
\includegraphics[scale=0.3]{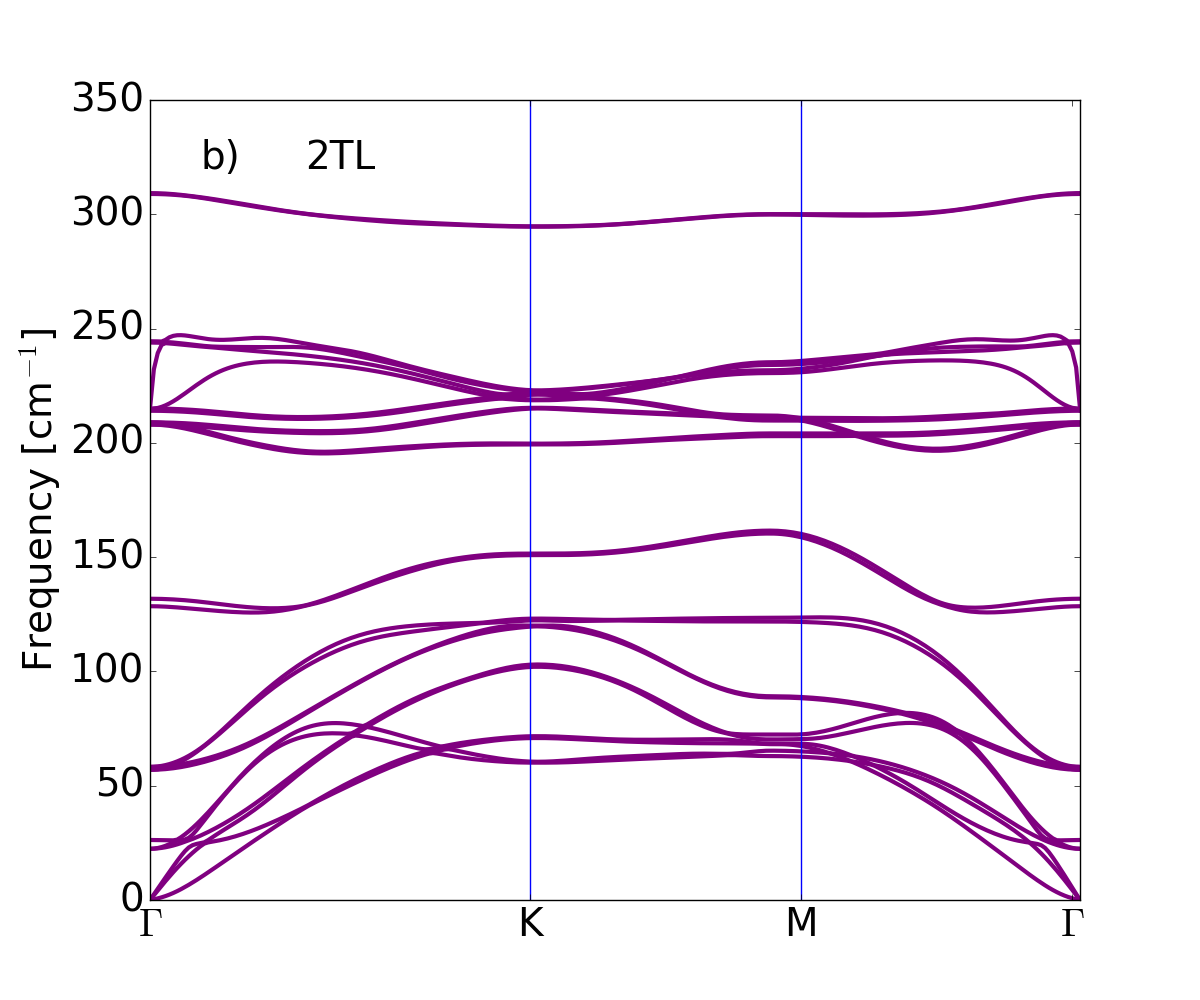}\\
\includegraphics[scale=0.3]{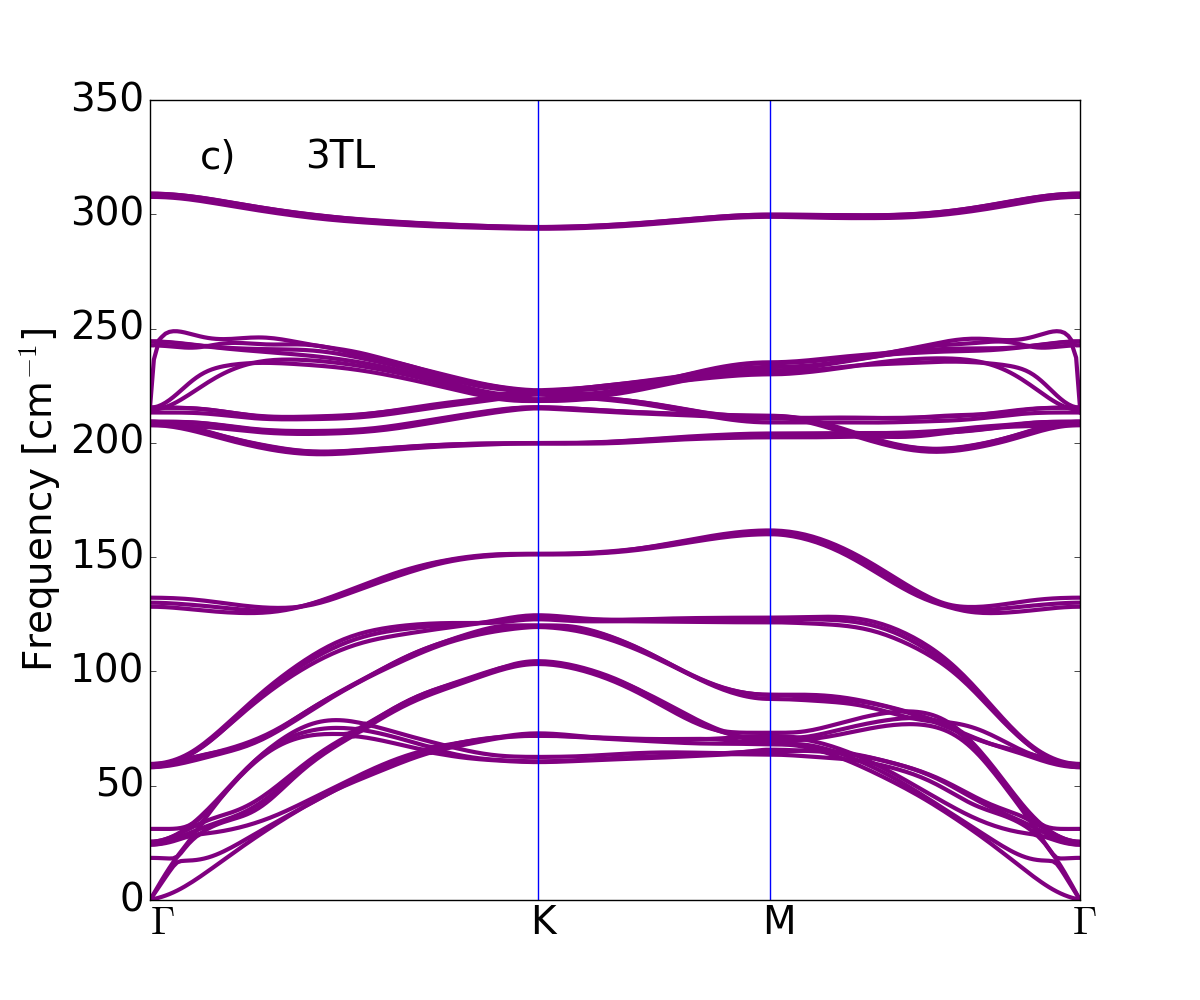}&
\includegraphics[scale=0.3]{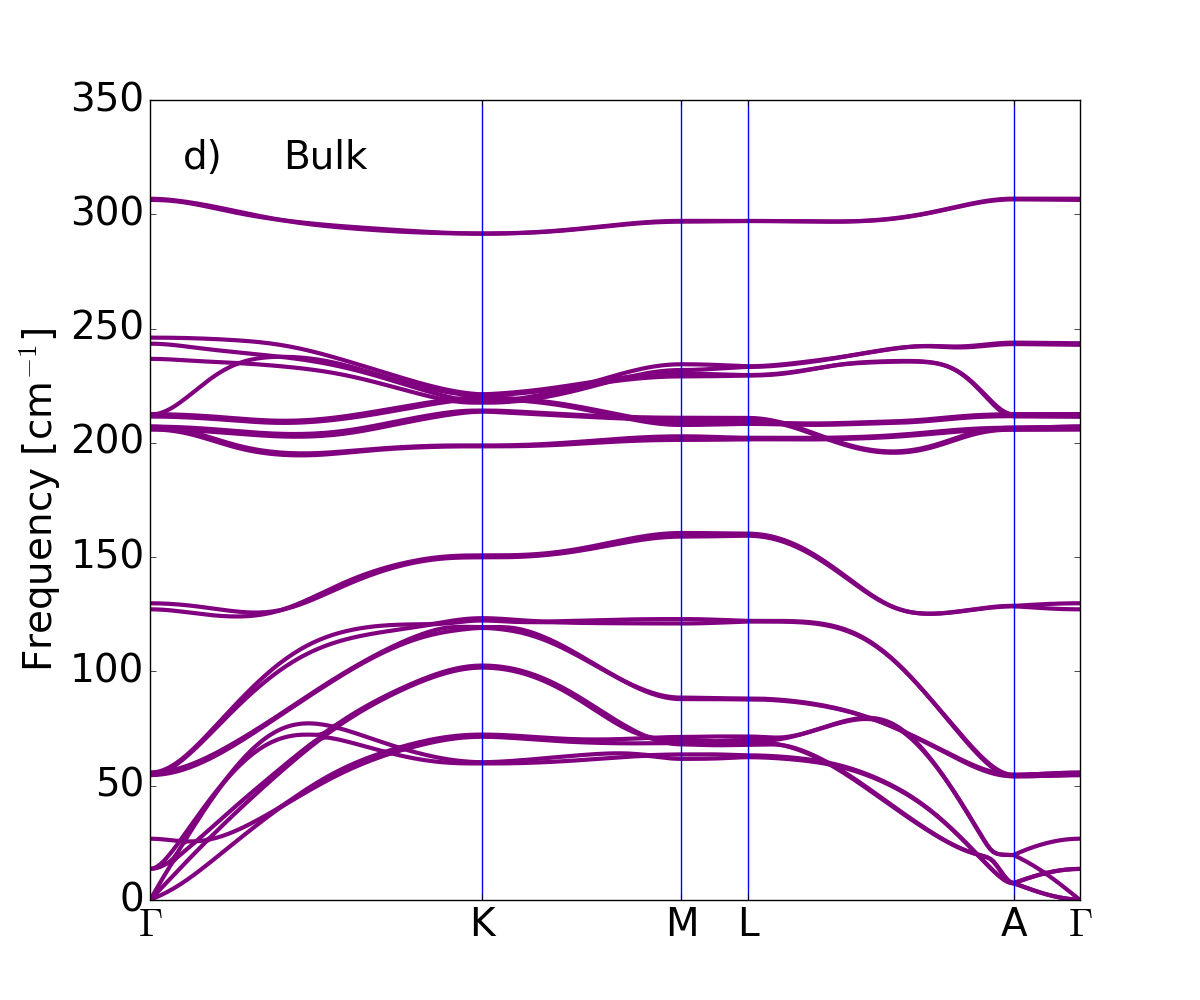}\\
\end{tabular}
\caption{\label{fig3}Phonon branches of 1TL, 2TL, 3TL, and bulk GaSe calculated with the LDA xc-functional. The branches are plotted versus high-symmetry lines in the 2D or 3D BZ (see caption of Fig.~\ref{fig2}). }
\end{figure*}

The dispersion of the phonon branches, obtained by means of the DFPT and the LDA xc-functional for LDA-optimized atomic equilibrium geometries, are plotted along the wave-vector path connecting $\Gamma$-$K$-$M$-$\Gamma$ high symmetry points in Fig.~\ref{fig3}, from single tetralayer to three tetralayers, and finally the bulk. 

\begin{figure}[h]
\includegraphics[scale=0.3]{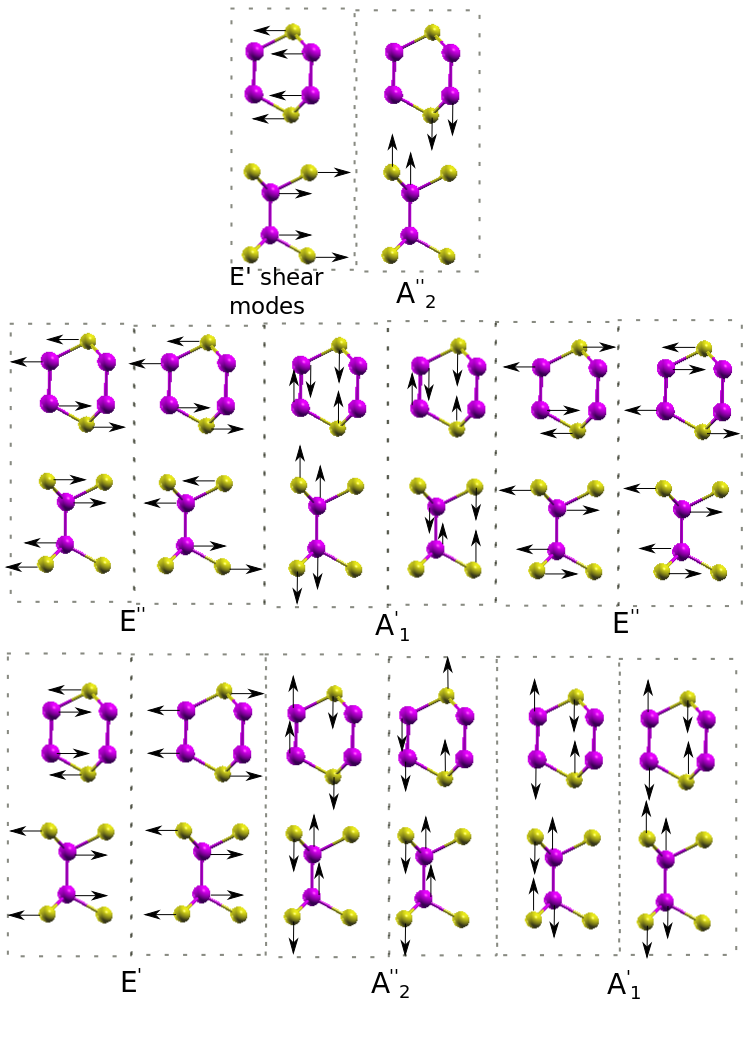}
\caption{\label{fig4} In plane $E'$ and $E''$, out-of-plane $A'_1$ and $A''_2$ normal modes of $\Gamma$ phonons in bulk $\varepsilon$-GaSe illustrated by atomic displacements of two tetralayers per unit cell. For each mode the antisymmetric and the symmetric displacements are shown, except for the first two modes ($E'$, $A''_2$ in the upper panels) where all the atoms belonging to one tetralayer move in the same direction. }
\end{figure}

The phonon modes of bulk $\varepsilon$-GaSe can be classified by means of the symmetry of the zone-center modes. The 24 $\Gamma$-modes can be decomposed into four groups $(A^{'}_1 \oplus A^{''}_2 \oplus E^{'} \oplus E^{''})$. Among these, $A^{''}_2 \oplus E^{'}$ are acoustic and $4 A^{'}_1 \oplus 3 A^{''}_2 \oplus 3 E^{'} \oplus 4 E^{''}$ are optical modes. Except for $A^{''}_2$ modes which are only infrared active, all the others are Raman active modes. Out of them, $E^{'}$ modes around 212 cm$^{-1}$ are also infrared active. The corresponding displacement patterns are displayed in Fig.~\ref{fig4}.

1TL and 3TL have the same space group as the bulk, with 12 symmetry operations, while 2TL has a $C_{3v}$ $(3m)$ symmetry with six symmetry operations. Since it is not centrosymmetric, all optical modes are simultaneously Raman- and IR-active \cite{Allakhverdiev.Baykara.ea:2006:MRB}.

The bulk phonon dispersion has three acoustic modes $A^{''}_2 \oplus E^{'}$. The $E^{'}$ mode vibrating in-plane (longitudinal and transverse acoustic) is two-fold degenerate and has a linear dispersion for vanishing wave vector. The $A^{''}_2$ mode is a flexural phonon mode (out-of-plane acoustic mode) which exhibits a peculiar quadratic dispersion near the zone center, typical of layered crystals \cite{Zabel2001} and which can be explained as a consequence of the point group symmetry \cite{Saito1998}. It is the easiest mode to be excited, because of its lowest frequency among all phonon modes and it may be then affected by numerical instabilities \cite{Zolyomi2014PRB}.
Here the three acoustic branches converge smoothly to zero frequency for vanishing wave vector, demonstrating that numerical instabilities have been safely controlled: no soft modes or negative energies appear in the phonon branches of bulk and GaSe nanosheets. We conclude  that not only the bulk $\varepsilon$-GaSe but also the layered systems are dynamically stable, at least, considering small 2D hexagonal unit cells. 

Table~\ref{tab3:GaSe_Omega} shows the bulk modes at $\Gamma$ together with their symmetry character. Modes of 1TL to 3TL are also listed, establishing the correspondence of few-tetralayer modes with bulk phonon symmetry representations. For each representation the number of phonon modes reflects the number of tetralayers. From 1TL to 3TL the phonon frequencies only vary of few cm$^{-1}$ because of the weak bonding between tetralayers.

GaSe is a polar material, hence the behavior of the longitudinal optical (LO) modes depends on dimensionality. The dipoles generated by the LO displacements interact with each other through long-range Coulomb interaction. This additional dipole-dipole interacting term causes the splitting between the LO and transverse optical (TO) phonon modes. In  bulk this splitting is independent of the phonon wavevector $\vec{q}$, in particular in the limit of $\vec{q} \rightarrow 0$, we find the LO mode at 242.6~cm$^{-1}$ and the TO mode at 212~cm$^{-1}$. In the isolated single tetralayer this splitting vanishes for $\vec{q} \rightarrow 0$ and in this limit the derivative of the LO phonon dispersion is discontinuous. Here the breakdown of the LO-TO splitting occurs at 213~cm$^{-1}$. Increasing the number of tetralayers the slope of the highest LO mode becomes large and the LO frequency smoothly approaches the bulk limit lifting the degeneracy \cite{Sohier2017NL}. 

Turning to the analysis of 1TL phonon band structures, two linear branches, related to in-plane vibrations, a transversal acoustic (TA) and a longitudinal acoustic (LA) one, can be seen in Fig.~\ref{fig3}(a). In addition, a flexural acoustic (ZA) branch is visible for out-of-plane vibrations. The two linear branches have large sound velocities of $\sim$2600~m/s (TA) and $\sim$4300~m/s (LA) in agreement with the experimental values~\cite{Abutalybov1995}. Approaching the $\Gamma$ point the ZA branch has the quadratic dispersion already discussed for the bulk case~\cite{Carrete.Li.ea:2016:MRL}. Immediately above the acoustic branches 
low-frequency $E'$ optical modes are present in the bulk (13.6 cm$^{-1}$), 2TL (22.4 cm$^{-1}$)  and 3TL (25.5 cm$^{-1}$). These modes are interlayer shear modes, typical of layered materials, corresponding to the sliding of two adjacent tetralayers (see Fig.~\ref{fig4} upper panel). No shear modes are obviously present in the 1TL case.

\subsection{\label{sec3d}Raman and IR spectra}

In the lower frequency range of bulk phonon band structure (see Fig.~\ref{fig3}(d) and Table~\ref{tab3:GaSe_Omega}) at the zone-center of the hexagonal BZ, a bunch of optical modes are found at 13.6~cm$^{-1}$ ($E^{'}$, the shear mode), 54.7 and 55.7~cm$^{-1}$ ($E^{''}$) as in-plane modes and 26.8~cm$^{-1}$ ($A^{''}_2$) as an out-of-plane mode. An in-plane mode in the lower frequency region is experimentally found around 59~cm$^{-1}$ (E'') in Raman spectrum \cite{Lei.Ge.ea:2013:NL}. In the intermediate frequency range we find two Raman active $A^{'}_1$ modes at 127.2 and 129.8~cm$^{-1}$ which correspond to a $A_1'$ dominant peak in the measured Raman spectra at 132~cm$^{-1}$ \cite{Lei.Ge.ea:2013:NL,Rahaman2018} - 134.4~cm$^{-1}$ \cite{Rodriguez.Mueller.ea:2014:JoVS&TB}. 
In the high energy region we find eight modes, two of them are $A_1'$ modes at 306.3 and 306.8 cm$^{-1}$, two Raman forbidden modes at 242.9 and 243.6~cm$^{-1}$ (LO modes) and 2 $E^{''}$, 2 $E^{'}$ modes around 206 and 212~cm$^{-1}$ (the latter is a TO mode) respectively. The consequent LO-TO splitting amounts to 30~cm$^{-1}$. Raman measurements reveal an intense peak at 305.2~\cite{Lei.Ge.ea:2013:NL} - 306~\cite{Rahaman2018} - 307.8~cm$^{-1}$~\cite{Rodriguez.Mueller.ea:2014:JoVS&TB} ($A_1'$) and an $E^{'}$ mode at 211~cm$^{-1}$~\cite{Rahaman2018} - 213~cm$^{-1}$~\cite{Rodriguez.Mueller.ea:2014:JoVS&TB}. A peak for $E^{''}$ mode around 208~cm$^{-1}$ appears in experimental Raman spectrum~\cite{Lei.Ge.ea:2013:NL} just for the bulk sample, while a peak at 230~cm$^{-1}$ is measured in single or double tetralayer samples accompained by a shift of about 3~cm$^{-1}$ to lower frequencies in thicker, few-tetralayer samples. The fact that we do not find this mode at such a frequency in few-tetralayer samples supports the hypothesis~\cite{Lei.Ge.ea:2013:NL} that this mode is due to the interaction of few-tetralayer samples with the substrate, i.e. it does not exist in the bulk.

\begin{figure}[h]
\includegraphics[scale=0.3]{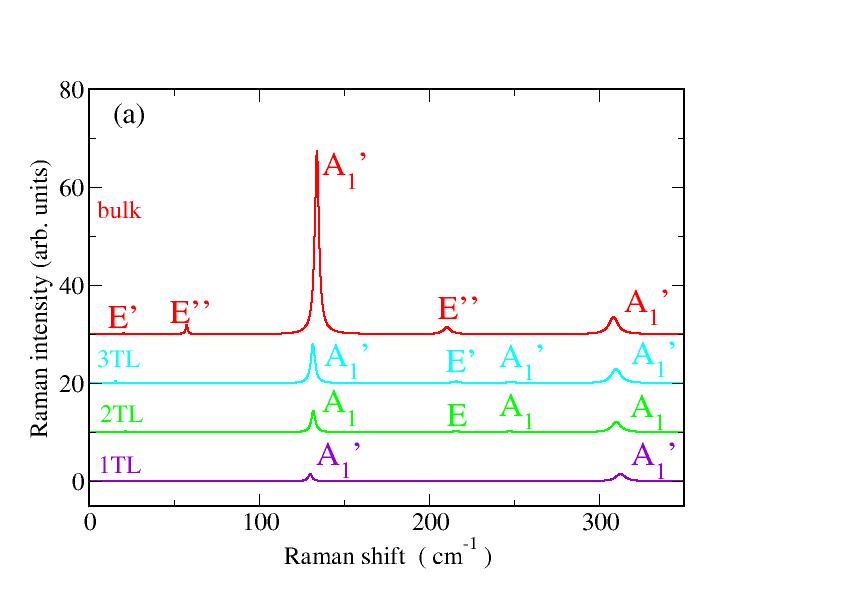}\\
\includegraphics[scale=0.3]{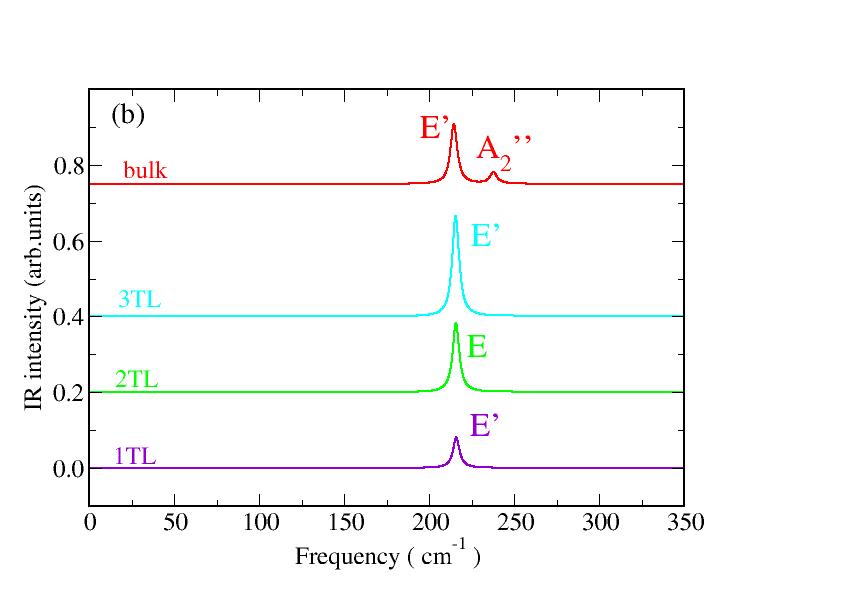}
\caption{\label{fig5}Calculated (a) Raman and (b) IR spectra using the frequencies and intensities from Table~\ref{tab3:GaSe_Omega}. The bulk (red), trilayer (blue), bilayer (green), and monolayer (purple) spectra are illustrated by different colors.}
\end{figure}

Calculated one-phonon Raman and IR spectra are displayed in Fig.~\ref{fig5} for bulk GaSe and varying the nanosheet thickness. Only zone-center phonons contribute to the spectra. 
In the 1TL case only the two $A_1'$ modes are clearly visible. As expected from group symmetry analysis, the twofold-degenerate $E'$ mode at about 213~cm$^{-1}$ shows a finite IR strength. 

Frequencies of Raman and IR modes at zone-center  of the same symmetry are grouped in blocks in Table~\ref{tab3:GaSe_Omega}. The average of the frequencies weighted by the Raman intensities leads to the position of the Raman peaks in Fig.~\ref{fig5}(a). 
Independently of the number of tetralayers two strong Raman peaks of $A_1/A_1'$ symmetry appear in the range of $306-309$~cm$^{-1}$ and $127-132$~cm$^{-1}$. 
The highest frequency mode is hardly layer-thickness-dependent, while the lowest one slightly decreases with decreasing  layer thickness. Both trends are confirmed by recent Raman measurements~\cite{Rahaman2018}. For what concerns the Raman intensities, in both cases the intensity increases with increasing  layer thickness. In particular, a really intense peak is visible in the bulk Raman spectrum  around 130~cm$^{-1}$. The Raman peak position  around $207-209$~cm$^{-1}$ related to $E''$ modes does hardly vary with the material thickness, in agreement with the fact that they are in-plane modes. The $A^{'}_1$ mode as well as the 
IR-active $A^{''}_2$ mode are out-of-plane vibrations and, therefore, more influenced by the interlayer interaction. 
Thereby, the highest $A^{''}_2$ mode frequency increases with reduced slab thickness, while its IR intensity decreases. 
Whereas the absolute peak positions, in particular for bulk, agree well with measurements, e.g. in Ref.~\cite{Rahaman2018} with deviations of about 1~cm$^{-1}$, and the thickness variation of the lowest $A^{'}_1$ mode are in agreement with the experimental findings, we report some discrepancies for the highest $A^{''}_2$ mode. 
We trace back the small differences between measurements \cite{Lei.Ge.ea:2013:NL,Rodriguez.Mueller.ea:2014:JoVS&TB,Rahaman2018} and our calculations to the presence  of a substrate in experimental setups, whereas only freestanding nanosheets are  theoretically investigated.

The IR spectra in Fig.~\ref{fig5}b are less complex compared to the Raman ones in Fig.~\ref{fig5}a. Only a strong in-plane $E'$ peak is visible in the range of $213-215$~cm$^{-1}$. In the bulk case this peak is slightly shifted toward $212$~cm$^{-1}$ close to the peak found in experimental GaSe spectra \cite{Giorgianni.Mondio.ea:1977:JPF}. Interestingly, we also predict a peak related to the out-of-plane $A_2''$ mode close to 243~cm$^{-1}$. However, the observability of the two peaks have to be carefully discussed in terms of ordinary or extraordinary light polarization.

\section{\label{sec4}Summary and Conclusions}

We have performed a theoretical study of 1TL, 2TL and 3TL GaSe nanosheets by means
of density functional theory. For the purpose of comparison the same computations have been done for the bulk $\varepsilon$-GaSe crystal polytype. Three different exchange-correlation functionals have been applied in the ground-state calculations. We found the typical underestimation (overestimation) of the chemical bonding using GGA-(LDA-)based functionals. We demonstrated that the LDA functional delivers the lattice constants in a sufficiently good quality. It can therefore be applied also for DFPT studies of the lattice vibrations. The investigations of the electronic structures clearly demonstrate that the GaSe nanosheets are indirect semiconductors because of a Mexican-hat dispersion of the uppermost valence band, while the bulk system is a direct semiconductor. Quasiparticle effects, in general, and confinement effects in the nanosheets increase the fundamental gaps toward values, which are comparable with experimental ones.

The lattice-vibrational properties are dominated by twofold degenerate in-plane $E$-type modes and non-degenerate out-of-plane $A$-type modes. The phonon band dispersion versus high-symmetry lines in the 2D hexagonal BZ shows that the few-tetralayer nanosheets are also dynamically stable and that the phonon band dispersion of 1TL, 2TL and 3TL exhibits the LO-TO splitting breakdown typical of polar two-dimensional systems. The frequencies, Raman intensities, and IR oscillator strengths allow the construction of Raman and IR spectra in qualitative but also quantitative agreement with available experimental data. This holds for the most intense peaks, in general, but also for the thickness dependence of their peak positions.

\begin{acknowledgments}
E.C.\ acknowledges support by the Programma per Giovani Ricercatori - 2014 ``Rita Levi Montalcini''.   O.P.   acknowledges  financial  support  from the EU MSCA HORIZON2020 project ’CoExAN’ (GA644076). CPU time was granted by CINECA HPC center. We thank T. Sohier and C. Attaccalite for useful discussions, A. Zappelli for the 
management of the computer cluster \emph{Rosa}.

\end{acknowledgments}

\bibliography{H:/Daten/TEXMEX/JabRef/literatur}

\newpage

\begin{table*}[h!] 
  \centering
  \begin{tabular}{|c c a|c c a|c c a|c c a|| c | c | c |} \hline      \multicolumn{12}{|c|}{Theory} &
  \multicolumn{3}{c|}{Exp.} \\ \hline 
    1 TL  & $\omega$ & &  2 TL  &  $\omega$ & &  3 TL    & $\omega$ &  & Bulk    & $\omega$ &  & 1-2TL & few TL & Bulk \\\hline
          &        &   & $E$   &  22.4  & I+R & $E^{''}$  &  24.2  & R &$E^{'}$  &  13.6    & I+R & & &  \\
          &        &   &       &        &     & $E^{'}$   &  25.5  &I+R&         &          &     & & &  \\\hline
          &        &   & $A_1$ &  26.3  & I+R & $A^{'}_1$ &  18.4  &R &$A^{''}_2$&  26.8    &  I  & & &  \\
          &        &   &       &        &     & $A^{''}_2$&  31.1  &I &          &          &     & & & \\\hline   
 $E^{''}$ &  55.2  & R & $E$   &  57.0  & I+R & $E^{''}$  &  58.1  &R &$E^{''}$  &  54.7    &  R  & 59 & 59& 59 \\
          &        &   & $E$   &  58.2  & I+R & $E^{'}$   &  59.2  &I+R&$E^{''}$ &  55.7    &  R  & & & \\
          &        &   &       &        &     & $E^{''}$  &  59.5  &R &          &          &     & & & \\\hline
$A^{'}_1$ & 128.8  & R & $A_1$ &  128.5 & I+R & $A^{'}_1$ & 128.3  &R &$A^{'}_1$ & 127.2    &  R  & 132 & 132 & 132  \\  	 
          &        &   & $A_1$ &  131.8 & I+R & $A^{''}_2$& 130.1  &I &$A^{'}_1$ & 129.8    &  R  & & &  \\
          &        &   &       &        &     & $A^{'}_1$ & 132.3  &R &          &          &     & & & \\\hline
 $E^{''}$ & 207.7  & R &  $E$  &  208.0 & I+R & $E^{''}$  & 207.8  &R &$E^{''}$  & 205.9    &  R  & N.O. & N.O. & 208   \\ 
          &        &   &  $E$  &  209.0 & I+R & $E^{'}$   & 209.2  &I+R&$E^{''}$ & 207.1    &  R  & & &  \\
          &        &   &       &        &     & $E^{''}$  & 209.2  &R &          &          &     & & &  \\\hline
  $E^{'}$ & 213.5  & I+R &$E$  &  214.0 & I+R & $E^{'}$   & 213.3  &I+R&$E^{'}$  & 211.5    &  I+R& & &  211$^{a}$  \\ 
          &        &     &$E$  &  215.1 & I+R & $E^{''}$  & 215.4  &R  &$E^{'}$  & 212.6    &  I+R& & & \\
          &        &     &     &        &     & $E^{'}$   & 215.4  &I+R&         &          &     & & & \\\hline   
$A^{''}_2$& 245.2  & I &$A_1$  & 243.9  & I+R & $A^{''}_2$& 242.7  &I &$A^{''}_2$& 242.9    &  I  & & &  \\  	          
          &        &   &$A_1$  & 244.5  & I+R & $A^{''}_2$& 244.3  &I &$A^{''}_2$& 243.6    &  I  & & &   \\
          &        &   &       &        &     & $A^{'}_1$ & 244.4  &R &          &          &     & & & \\\hline
$A^{'}_1$ & 309.7  & R &$A_1$  &  308.9 & I+R & $A^{'}_1$ & 307.6  &R &$A^{'}_1$ & 306.3    &  R & 303.4& 303.4 & 305.2 \\ 	 
          &        &   &$A_1$  &  309.3 & I+R & $A^{'}_1$ & 309.0  &R &$A^{'}_1$ & 306.8    &  R & & & 306$^{a}$  \\
          &        &   &       &        &     & $A^{''}_2$& 309.0  &I &          &          &    & & &  \\\hline
  \end{tabular}
\caption{Calculated optical modes of GaSe nanosheets from 1TL to 3TL and bulk $\varepsilon$-GaSe at zone center characterized by their corresponding irreducible representation. Frequencies are in cm$^{-1}$. Raman active modes are labeled with "R", IR active modes with "I", and experimentally not observed modes with "N.O.". $E$ modes are twofold degenerate, while $A$ modes are not. Experimental values are extracted from Ref.~\cite{Lei.Ge.ea:2013:NL} and $^{a}$~Ref.\cite{Rahaman2018}}
\label{tab3:GaSe_Omega}
\end{table*}

\end{document}